%% file: fiability.tex
\documentclass{article}

\def\Section#1{\section{#1}}
\def\SubSection#1{\subsection{#1}}
\def\SubSubSection#1{\subsubsection{#1}}

\usepackage{fiability-macros}
\usepackage{RR}

\def\figsize{12cm}

\begin{document}

\RRdate{November 2006}
\RRNo{1825}
\RRetitle{Increasing Data Resilience of Mobile Devices with a Collaborative Backup Service}
\RRtitle{Fiabilit\'e des sauvegardes dans un service de sauvegarde collaboratif pour terminaux mobiles}
\RRauthor{Damien Martin-Guillerez, Michel Ban\^atre, Paul Couderc\thanks{\tt\{dmartin,banatre,pcouderc\}@irisa.fr}
}
\authorhead{Martin-Guillerez \& Ban\^atre \& Couderc}
\titlehead{Increasing Data Resilience of Mobile Devices}
\RRprojet{ACES}
\RRtheme{\THCom}
\RRabstract{\input{fiability-abstract}}
\RRresume{\input{fiability-resume}}
\RRkeyword{Data resilience, mobile computing, collaboration, backup, sensor networks, mobile ad~hoc networks, data MULEs.}
\RRmotcle{Tol\'erance aux fautes, informatique mobile, syst\`eme collaboratif, sauvegarde, r\'eseaux de capteurs, r\'eseaux mobile ad~hoc}
\RRnote{This work is partially funded by the ACI SI MoSAIC and the ReSIST Network of Excellence}
\URRennes
\makeRR

\input{fiability-text}

\bibliographystyle{abbrv}
\bibliography{fiability}

\end{document}

%% file: fiability-abstract.tex
Whoever has had his cell phone stolen knows how frustrating it is
to be unable to get his contact list back. To avoid data loss when
losing or destroying a mobile device like a PDA or a cell phone, data
is usually backed-up to a fixed station. However, in the time
between the last backup and the failure, important data can have
been produced and then lost.

To handle this issue, we propose a transparent
collaborative backup system. Indeed, by saving data
on other mobile devices between two connections to a global
infrastructure, we can resist to such scenarios.

In this paper, after a general description of such a system, we
present a way to replicate data on mobile devices to attain a
prerequired resilience for the backup.

%% file: fiability-resume.tex
Quiconque \`a d\'ej\`a perdu son t\'el\'ephone portable sait qu'outre
la perte mat\'erielle, la pertes de la liste des contacts est tr\`es
g\'enante. Pour \'eviter toute perte de donn\'ees lors de la destruction
ou la perte d'un appareil mobile tel qu'un PDA ou un t\'el\'ephone portable,
les donn\'ees sont habituellement sauvegard\'ees sur une station fixe.
Cependant, les donn\'ees acquises depuis la derni\`ere sauvegarde seront
d\'efinitivement perdues.

Pour proteger ces donn\'ees, nous proposons d'utiliser un syst\`eme de
sauvegarde collaborative. En effet, sauvegarder les donn\'ees importantes
sur les terminaux voisins via un dispositif de communication sans-fils
permettrait de palier ce probl\`eme.

%% file: fiability-text.tex
\sect{Introduction}{intro}

The use of mobile computers, such as laptops, PDAs, mobile phones or
digital cameras, has increased amazingly during past years. Thus, the
production of sensible data on such device has also increased. The
loss of such data can have painful consequence for users : loss of
phone numbers, loss of meeting dates, or deletion of important notes
or pictures.

To reduce data loss, those devices usely have a synchronization-like
mechanism, which main issue is that you need to be near your computer
bringing up time periods during which device failure means
irreversible data loss. For example, if you take a note on your PDA
during a meeting and your PDA get lost, stolen or broken on your way
back, then the note is definitely lost.

However, more and more mobile devices come with wireless connectivity
like IEEE 802.11 or Bluetooth. Using neighbor devices to save data
right after its production can decrease data loss by restoring data
either from a global-scale network like the Internet or directly from
a backup device. Saving automatically on a global-scale network seems
to be a viable assumption because of the growing number of wireless
access to the Internet. Nevertheless, the required infrastructure for
this kind of access is expensive (e.g. GPRS, UMTS). In such a
situation, the use of neighbor peers to backup sensible data is a way
to decrease the cost of the backup.

We aim at designing and implementing a transparent collaborative
backup service for mobile devices~\cite{Mosaic2004a}. Such a service
differs from existing works and thus needs to meet specific requirements
we outline in \sref{design}. Then, we analyze several issues
specific to mobile device data and replication in
\sref{issues}. Afterwards, we present a way to order replicas in
that system in \sref{reliability} and ideas for backup terminals to
manage replicas in \sref{deletion}. Finally, after outlining works
that are still pending in \sref{future}, we present existing systems
in \sref{related} and conclude in \sref{conclu}.

\sect{Design overview}{design}

Our main purpose is to design an efficient backup system called
MoSAIC~\cite{Mosaic2004} that can handle high mobility, which means
that it needs to handle two scenarios:
\begin{itemize}
  \item[-] When connected to a global network like the Internet, the
  system must use this opportunity to save data on a resilient server.
  \item[-] When disconnected from the global network, it must use
  neighbors to backup selected data (i.e. data of higher
  importance).
\end{itemize}

Also, depending on data production (e.g. production rate, data
importance), the system should adapt the level of replication. We
especially want fair use of the system to avoid useless resource
consumption. Moreover, the system needs to be protected against
egoistic participants that backup but do not provide resources to
others.

Furthermore, the system should avoid useless energy consumption. As
the system should work on mobile system, energy and other resource are
quite scarce and should be used wisely.

We want the system to be as implicit for the user as
possible. That means:
\begin{itemize}
  \item[-] very few actions are required from the user when performing the
    backup or the recovery (i.e. the backup needs to be a complete one
    and easy to restore),
  \item[-] no prior trust relationship with other peers is required,
  \item[-] no extra hardware is required.
\end{itemize}

As shown in \fref{propagation-scheme}, a client terminal can either
backup its data to another terminal (the backup peer) or to an
Internet server. Data can be transfered from the backup peer to the
Internet server. The client terminal can then restore its data either
from a backup peer or from the Internet server. We do not consider to
propagate backup through peers due to two reasons:
\begin{itemize}
  \item[-] Copy of backup through terminals costs energy and others
  resources. Just propagating replica with deletion of the original
  one costs in communication resources (e.g. energy and time) and does
  not improve backup reliability.
  \item[-] Only the owner of the data can know when it's necessary to
  start a replication. A replication issued by a backup terminal has
  a high probability to be useless.
\end{itemize}

\fig[\figsize]{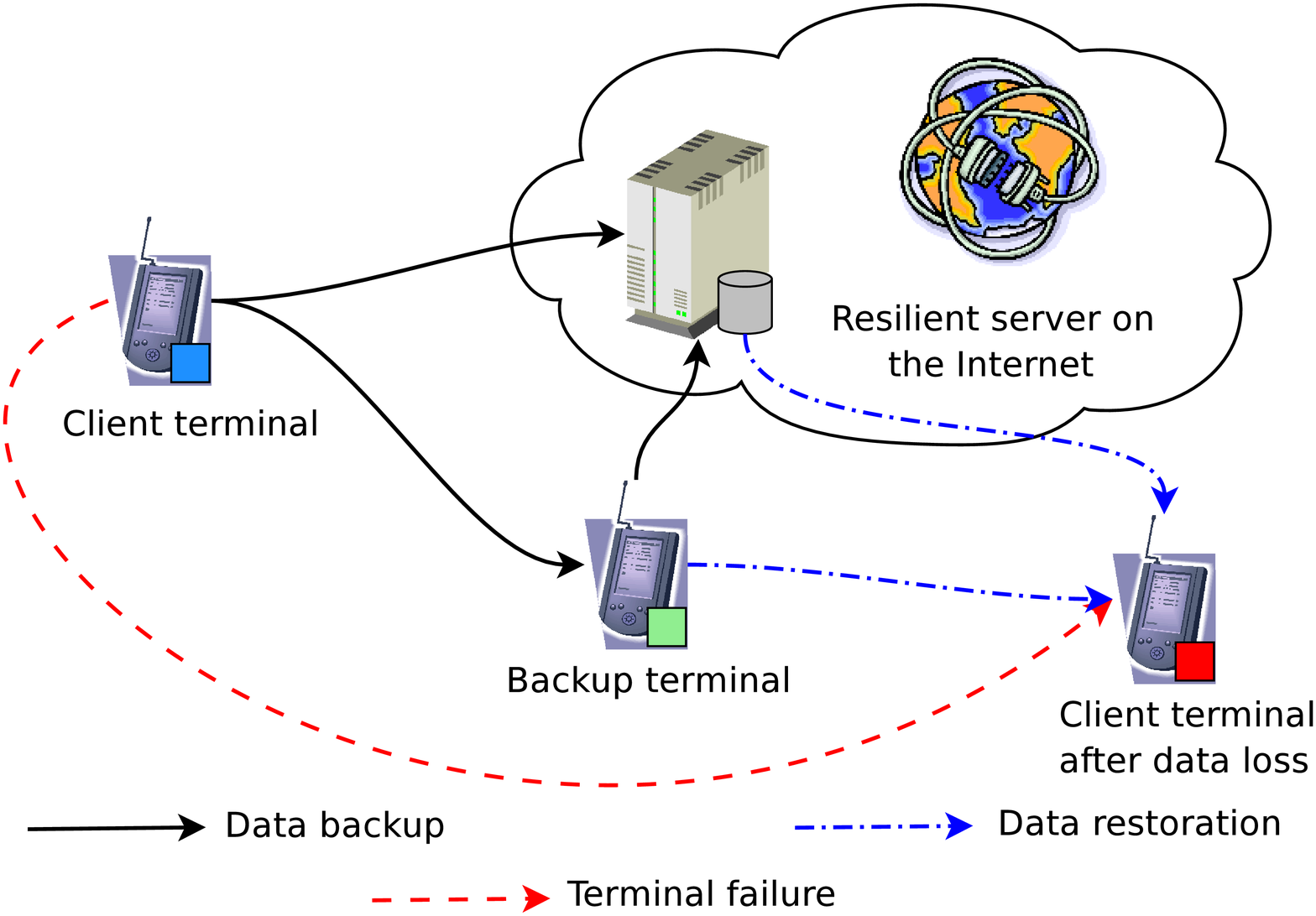}{Considered backup scenarios.}

That scheme also fits well for data MULE~\cite{Shah2003}
networks. Data MULEs are mobile wireless terminals or sensors that
carry data from a location to another by the mobility of its
carrier. For example, Burrell et al.~\cite{Burrell2004} propose to use
a sensor in a shovel or other tools that collect data from sensors in
the vineyard so that the computer at the farm will be able to analyze
data brought back by the movement of the farmer.

In the same way, cattle health can be monitored using sensors that
transmit data to a base station. Either health data, like temperature,
or alerts can be issued by sensors. Those data, especially alerts,
need to reach the base station even if the sensor fails. Using
the proposed system can help those data to reach the base station.
Sensors on birds that keep tracks of encounters can be used to monitor
epidemics. During the encounters, the sensors can also use neighbors to
save their data and optimize the reading by the observer (he just has
to read one bird's sensor instead of all sensors).

Besides the two previous architecture, Ban\^atre et
al.~\cite{Banatre1998} propose to use collaborative robots to realize
tasks without a centralized brain. In that system, data
gathering and transmission are key points. In that system some
information should be backed-up to a ``local brain'' (i.e. a reliable
storage close to the robots) or a global server. Then the MoSAIC
scheme can be applied to increase data availability or reduce the need
for a global wireless coverage area.


\sect{Data issues}{issues}

\subsect{Mobile device data}{data}

In this section, we look at produced data on classical mobile
devices (i.e. PDAs and mobile phones) and at their attributes to
understand their specific issues.


The first data attribute is, obviously, the size. The size depends
highly on the types of data: they go from less than 200 bytes for SMS
or a schedule entry, to hundred of megabytes for video captures. The second
attribute is the production method. For instance, if a note can be
created, updated or deleted, pictures are generally only created or
deleted on those devices. Another attribute is the importance of data,
which can be from high for notes taken during a meeting to very low
for holiday pictures. Dependencies are also important: an e-mail can
be useless until you have all preceding e-mails in a discussion
thread. When a data item depends on preceding data like in a
discussion thread, we call that dependency a {\it temporal
dependency} contrary to a {\it spatial dependency} where a data item
$D$ depends on several others that can depend on $D$. Finally, a last
attribute that interests us is the life time of data. Actually, some
data like schedule entry become less important when the time of the
event has passed, even if it may be still important to save it.

In the case of data MULEs, data items are generally small entries
(current temperature) or track of past events (encounters for epidemic
monitoring). I.e. we can consider that those data items are small
entries with potential temporal dependencies.

So, mobile device data can be categorized by size, production method
(creation only, read/write, append only), dependencies (temporal and
spatial), life time and importance.

The size strongly affects the backup system in order to:
\begin{itemize}
  \item[-] Resist to mobility or network problems during a
  transmission. The capacity of a transmission depends mainly on the
  bandwidth and the connection time. The MTU (Max Transfer Unit) can
  also be important. While bandwidth and MTU are generally easy to
  know, the connection time depends more on mobility.
  \item[-] Avoid to monopolize one terminal memory. Memory consumption
  is a critical aspect in a mobile backup system. In the same way,
  when backing-up a data item on a mobile terminal, the size of the item
  affects the length of the transmission and thus the energy consumed
  by the backup terminal. Therefore, deletion of replicas can be
  needed to free some space on terminals. It can be decided depending
  on the size of the replicas, on the arrival of a new version, on the
  number of replicas, etc...
\end{itemize}

On the other hand, production method affects the part of data that
needs to be saved (i.e. the entire file or the new entry, etc...) and
the dependency (e.g. when backing-up just an entry that depends on
other). Moreover, dependencies affect the integrity of the backup and
thus needs some version tracking presented in
\sref{version}. Finally, we affect a priority to each data item
relatively to its importance and try to save data with highest
priority first (\sref{reliability}).

\subsect{Dispersion of replicas}{dispersion}

Since data size can be quite huge, there is a need for fragmentation
of files. Moreover, the high probability of a terminal failure to
restore a replica creates a need for a flexible replication
scheme.  Courtes et al.~\cite{Courtes2006} have already looked at
methods for redundancy and compression in that system. First, we
consider that all the data items that have spatial dependencies are
agglomerated into one data item (the priority of the new item is the
highest of the agglomerated items) so that the only dependencies we
consider are the temporal ones.

Then, we consider the $(n,k)$ replication scheme (as in Rabin's
information dispersal algorithm~\cite{Rabin1989}) that fragments the
data into $n$ fragments where only $k$ are required to reconstruct the
data. We also consider delta-compression which saves only the
differences between an old version and a new version of the same
file. While the $(n,k)$ replication scheme creates loosely spatial
dependencies, delta-compression creates strict temporal
dependencies. Simple replication is just a $(n,k)$ replication scheme
with $k = 1$. Replication when delta-compressing is made on
generated delta.

So, we now consider the following format for every data item to save:
\begin{itemize}
  \item[-] $n$ fragments.
  \item[-] Only $k$ fragments are required to reconstruct the data item.
  \item[-] The data item can have temporal dependencies on some other
  data (and then the priority of old data should be increased if the
  priority of the new data item is higher).
\end{itemize}

\subsect{Versions tracking}{version}

\fig[\figsize]{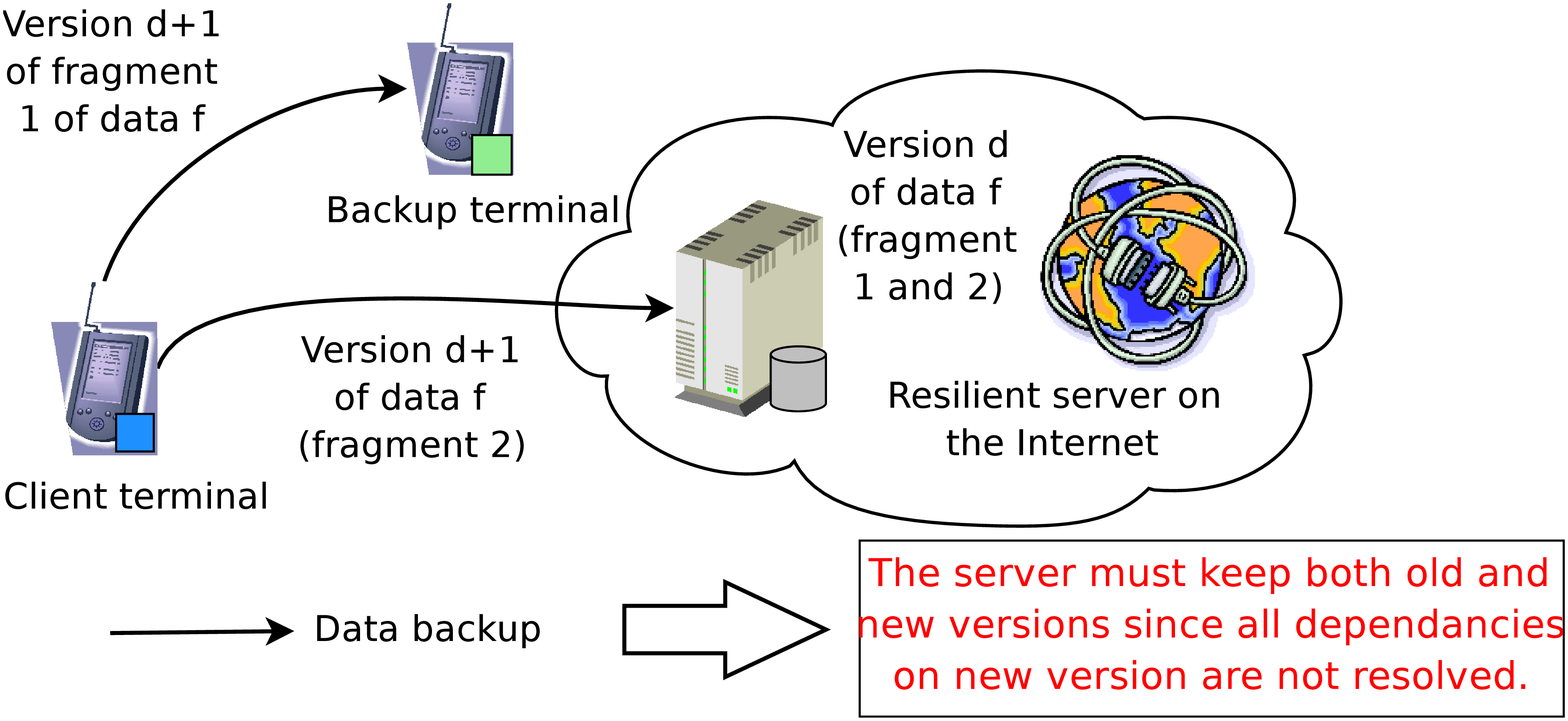}{Dependencies can prevent us to free some memory
  space when a new version of a data item arrives.}

Given considered propagation and dissemination schemes, some issues
can appear regarding arrival of new version of a data item to
backup. Firstly, in presence of dependencies, the old version of a
data should be kept until all dependencies of the new version have
been backed-up to the resilient server exhibited in \fref{dependency}).

\fig[\figsize]{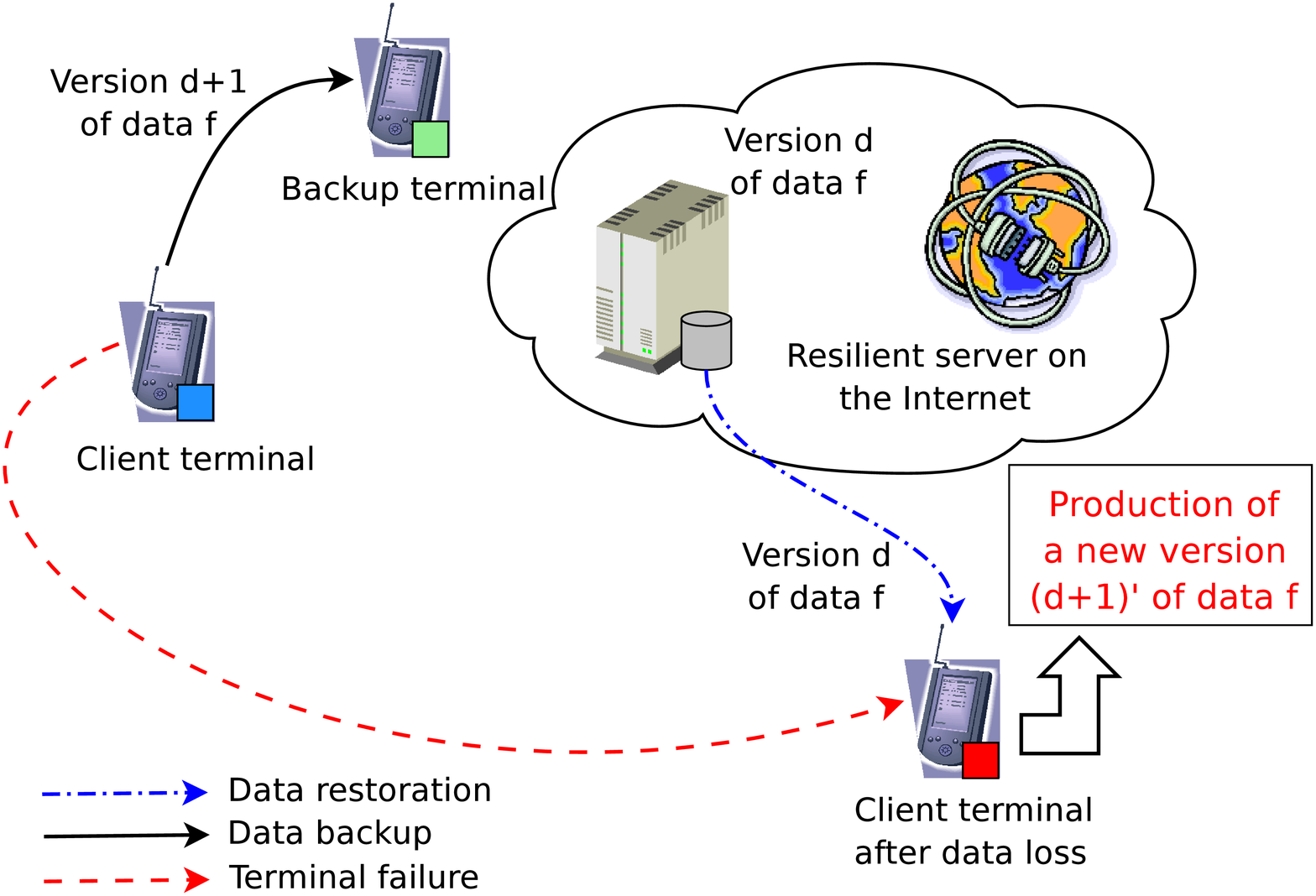}{Conflict appearing during a restoration.}

Conflicts may appear in our system. As a matter of fact,
when you backup the data of a mobile device on a fixed station, no
conflict can occurs since all versions of the same data item are on
the same device (the mobile terminal). But, considering our
propagation scheme, a conflict may appear (see \fref{conflict}) when a
data item is backed-up on another mobile device and an old version of
this same data item is located on the Internet server: if a failure
occurs, the client may restore the old version from the server and
work on it, generating a conflict with the version backed-up on the
mobile device. When facing such a situation, our system must use
conflict resolution mechanisms such as in Coda~\cite{Kumar1995} or
Bayou~\cite{Terry1995}.

Regarding those issues, the system must keep track of
replicas' versions.

\sect{Estimating backup reliability}{reliability}

In this section, we look at how to estimate in real time the
probability of a data item to be correctly restored and how we can use
this estimation to order backups.

\subsect{Reliability estimation}{estimation}

For the moment, let just consider the $(n,k)$ replication scheme. Let
$\P_i$ be the probability of getting back the replica $i$ and $\P^l_i$
the probability of being able to get back $l$ replicas between the
first $i$ ones. Then we can infer from \fref{proof}):
\begin{equation}
\P^l_i = (1 - \P_i) \cdot \P^l_{i-1} + \P_i \cdot \P^{l-1}_{i-1} 
\label{eqresiliency}
\end{equation}
\fig[\figsize]{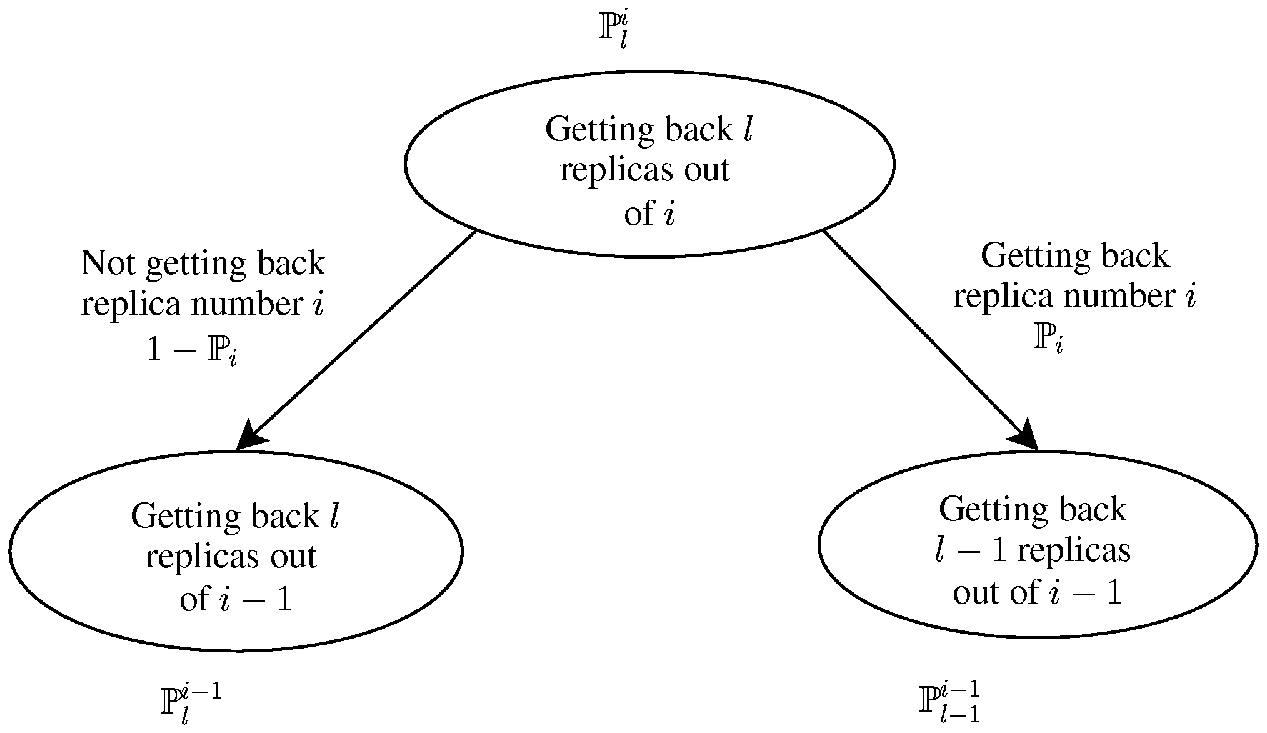}{Graphical proof of equation (\ref{eqresiliency}).}

In particular,
\begin{eqnarray}
\P^k_k & = & \prod_{l=1}^k \P_l \\
\P^l_i & = & 0\ \text{ if }\ i < l \\
\P^0_i & = & 1
\end{eqnarray}

When backing-up an additional replica, we can estimate the
influence on the probability of getting back the entire data
item. That is correct, of course, only if we save each replica on a
different terminal which means that all $\P_i$ are independent. We
can handle the case of two replicas being backed-up on the same
terminal by considering that they will have the same probability
$\P_i$, which is a viable assumption since they use the same {\it
transmission canal}. Thus, if we save $m$ replicas onto
the same terminal at the same time, the equation (\ref{eqresiliency})
becomes:
\begin{equation}
\P^l_{i+m} = (1 - \P_{i+1}) \cdot \P^l_i + \P_{i+1} \cdot \P^{l-m}_i 
\label{eqdependancy}
\end{equation}
We consider that saving another replica later on an already used
terminal is an independent event because too much time generally happens
between two encounters of the same terminal.

The last things that we must take into account are the temporal
dependencies. The probability of correctly restoring a new data item
depending on old ones is the probability of restoring the new data
item multiplied by the probability of restoring the old data items.

Considering all those points, we can estimate the probability of a correct
recovery during the backup itself. 

\subsect{Priority and replica scheduling algorithm}{priority}

We said in \sref{data} that each data item is associated with
a priority. That priority is supposed to be established by prior
mechanisms (user intervention for instance) and is given as a desired
backup resilience (e.g. a probability). We can classify data to be
backed-up using a queue ordered by the priority of the data item minus
the computed probability of successful backup.

$F_p$ is a priority queue of data to save. The $priority$ field of
a data structure is the priority affected to a data item. The
\aref{meeting} shows a general algorithm to order data packet to
save. First, if we have data to save, we try use the terminal until it
becomes unreachable. We pull off the queue the first data item which
can be saved on terminal $t$. Then we save the next packet (index $i$)
and recompute the probability of a successful backup. If the
probability is not high enough then the data item is re-enqueued in
$F_p$.

\begin{algo}{When meeting another terminal}{meeting}
 \algname{OnMeeting}{$t$}
 \begin{algtab}
   \algwhile{\algcall{Reachable}{$t$} \algand \textsc{DataToSave}}
      $d \leftarrow$ \algcall{Pull}{$F_p$, \textsc{CanSave}, $t$} \\
      \algifthen{\algnot \algcall{Exists}{$d$}}{\algbreak}
      $l \leftarrow$ \algcall{NextPacket}{$d$}\\
      $p \leftarrow$ \algcall{Save}{$t$, \algcall{GetPacket}{$d$, $i$}} \\
      $proba \leftarrow$ \algcall{RecomputeP}{$d$, $p$, $t$, $i$} \\
      \algif{$proba < d.priority$}
        \algcall{Push}{$F_p$, $d$, $d.priority - proba$}\\
      \algend
   \algend
 \end{algtab}
\end{algo}

\textsc{RecomputeP} in \aref{meeting} can be done using equations
(\ref{eqresiliency}) and (\ref{eqdependancy}). It needs to keep $k$
entries (to keep $(\P^l_i)_{1 \leq i \leq k}$) and to recompute them
each time a new packet is added (and thus needs $k$ operations).
Therefore, we have a realistic real-time algorithm to order replicate.

\sect{Replica management by backup terminals}{deletion}

Using neighbors wireless appliances for data backups consumes
resources as mentioned in \sref{design}. In this section, we
concentrate on memory usage. If we consider assigning a certain amount
of memory to the system, free space can become a problem after a
certain time. Firstly, an appliance can need more memory to perform its
tasks. Secondly, replicas more important than those stored on the
appliance can be refused due to a lack of memory. So we must see
which are the criteria to manage replicas on backup terminals.

\subsect{Detecting useless replicas}{useless-replicas}

A replica becomes useless either when it has been saved on the
destination server or when it has been outdated. A replica being
outdated means that either its data are no more useful (like
one-month old temperatures if we just want less than one-week old
ones) or that it has been updated by new data (like a schedule entry
being replaced).

A terminal can know when a replica has been saved on the resilient
server or has been updated when either:
\begin{itemize}
  \item[1.] the backup has been performed by the terminal,
  \item[2.] the owner has notified the terminal,
  \item[3.] the server has notified the terminal,
  \item[4.] a notification has been issued by other terminal.
\end{itemize}

While cases 1, 2 and 3 can be done when interacting with either the
terminal or the server, the case 4 needs propagation and thus can
waste communication resources.

The lifetime of a replica can be given by the owner when doing the
backup. Besides deleting replicas after a certain time, we can easily
add messages to say that a replica is no longer needed during other
transmission but an efficient protocol has to be designed to do
it. Moreover, we must look at the cost of notification propagation and
the related security problems.

\subsect{Criteria to free memory when needed}{lazy-deletion}

After a long disconnection time, memory usage can become a problem
either for our system or for the classical terminal usages even if
outdated or backed-up replicas are deleted. To handle this issue,
some replicas must be deleted based on partial informations. When
deleting such a replica, several criteria should be taken into
account:
\begin{itemize}
  \item[-] the age of the replica can help the terminal to estimate if
  it has been backed-up, updated or if its has a good chance
  to be no longer relevant.
  The time period between connections of the owner of the replica and
  the mean time before a replica reaches the Internet in the system
  can be useful to estimate the life-time.
  \item[-] the backup resilience and its importance can be used to
  select the less relevant replicas. When a replica has reached a
  high resilience compared to its importance, deleting it can be
  painless. Of course, we want to be fair towards each user and prevent
  a user to declare that all his data is important (we can also include
  more trusted terminals to the comparison). 
  \item[-] the data size is also important: the more space we get
  back, the better. It is more efficient to delete a lot of data items
  from the same user than deleting data of several users (notably
  when they are data of an untrusted user). Furthermore, a replica can
  have a lot of dependencies that can be deleted at the same time than
  the replica.
  \item[-] It is also possible, especially in the case of data MULEs,
  to merge replicas between terminals to free memory from one to
  other. If we look at data from epidemic tracking sensor
  networks, tracks of animal encounters can be merged into a single
  one on a single sensor.
\end{itemize}

We have seen a lot of criteria that can be used to determine which
replica to delete or to merge. Many of them require some information
and a specific communication protocol. Some security mechanisms are
also needed to prevent either automatic deletion by a terminal or lies
on the importance of their data.

\sect{Future works}{future}

We have seen a general design of the system and an algorithm to order
backups. We will now look at open problems and especially those on
which our future works are scheduled..

We have seen several requirements in \sref{design}. Firstly, the
system must be user-transparent. Thus, in the proposed algorithms,
priority of the backups must be determined by the system itself using
knowledge on the data (and can thus can depends on the
context of the user).

Secondly, the system should not rely on prior relationship. I.e.  the
system needs confidentiality techniques and incentives. Indeed, in the
MoSAIC system, each terminal does not known each other a priori and
thus entrusting backups to a terminal means 1) protecting data from
beeing reads by backup peers and 2) beeing able to entrust the backup
process to the peer. Of course, this does not apply for data MULEs or
collaborative robots because all the peers belong to the same
system. The only needed protection might be encryption against
outsiders listening to the communications.

Thirdly, the system should not rely on specific hardware but on
classical wireless interfaces but without interference with classical
use of those interfaces. In addition, the network layer should take
into account the high mobility and the energy consumption.

One main pending issue is the estimation of the probability of one
packet being correctly restored ($\P_i$). The main parameter is the
reliability of the device itself. An evaluation of this reliability
can be given by incentives. Other parameters can be battery lifetime,
terminal context (during holidays, it has less chance to get in touch
with the Internet than during workdays) and available memory. Thoses
parameters will be different for data MULEs and robot networks.

We have evoked resource management related issues such as decreasing
memory and energy consumption. We have also talked about deletion of
replicas to free memory for other ones. More works are needed in
order to know how to select replicas that can be deleted and how they
affect the efficiency of the backup system. Besides, for data MULEs and
collaborative robots, all appliances can read the data (there is no
need for confidentiality technics) and thus get a better understanding
on the way to reduce memory usage by doing data aggregation (like
in~\cite{Srivastava2006}) and backup reconciliation~\cite{Minsky2002}.

The delivery of data on the Internet by a contributor should be fast
and light. That is to say that we need to cleverly design the delivery
protocol to reduce the traffic between contributors and Internet
servers. Besides, for colaborative robots, we can envision a ``local
brain'' which see only part of the informations located on the
Internet server. The same thing can be considered for classical mobile
devices where we can envision the presence of
Infostations~\cite{Frenkiel2000}. In mobile sensor networks, sensor
readers are the same as Infostations or the ``local brain'': they
generally have little knowledge of the whole storage but are reliable
storage in themself. The main point with thoses ``local brains'' is
that the devices should backup data to them but avoid traffic usage
with transmitting already saved backups.

\sect{Related works}{related}

Increasing data resilience is usually done through hardware
replication~\cite{Patterson1988}. In network file systems, replication
of data can be realized using several data
server~\cite{Satyanarayanan1990}. Recently peer-to-peer file systems
have used replication to increase data
availability~\cite{Kubiatowicz2000} and have paved the way for
collaborative backup services~\cite{Batten2001,Cooley2004}.

In a mobile context, Rumor~\cite{Guy1998} and Roam~\cite{Ratner1999} use
peers to replicate data for high availability but can hardly handle
high mobility due to the clusterization of the peers. Indeed, when a
peer moves from a cluster to another, data replication between clusters
is needed. In fact, Rumor and Roam are not designed for backup
recovery but for high availability and data sharing. Moreover,
neither Rumor nor Roam exploits opportunistic replication on random
mobile peer.

AdHocFS~\cite{Boulkenafed2003} and Segank~\cite{Sobti2004} provide the
same facilities as Rumor and Roam. They are file systems that
focus on high availability and data sharing. AdHocFS transposes
peer-to-peer file systems' paradigms to ad hoc networks and Segank
concentrates on one person's devices (either mobile or fixed) to get a
uniform file system. Therefore, neither AdHocFS nor Segank gives
support for high mobility.

FlashBack~\cite{Loo2003} is a backup system for mobile device that
can handle quite efficiently data loss or even device failure like
destruction or robbery. However, FlashBack uses devices of a Personal
Area Network (PAN) to manage the backups. Hence, FlashBack is designed for
people with several wireless mobile devices on them.

On the contrary, we aim at creating a backup system that can be used
on wireless mobile devices without other prerequisites. We especially
want to handle high mobility and to get advantage of randomly encountered
peers with no prior trust relationship (contrary to Segank or AdHocFS).

\sect{Conclusion}{conclu}

Existing backup systems for mobile device usually rely on
pre-established trust between all participants and very light
mobility. We have presented a general design for a backup system that
can handle high mobility and does not rely on pre-established
relationship. We have outlined several issues concerning this system
and presented an algorithm to order replicas.

Issues regarding incentives, confidentiality, high mobility and
resource management are still to be resolved. In the near future, we
will concentrate on resource management, especially strategies for
replicas replacement in the special case of mobile sensor networks and
collaborative robots.

%% file: fiability.bbl
\begin{thebibliography}{10}

\bibitem{Banatre1998}
M.~Banâtre and F.~Weis.
\newblock Système d'information spontané (sis) : Problématique et premier
  éléments de solutions.
\newblock Technical Report 1222, IRISA, Dec. 1998.

\bibitem{Batten2001}
C.~Batten, K.~Barr, A.~Saraf, and S.~Treptin.
\newblock {pStore: A Secure Peer-to-peer Backup System}.
\newblock Technical Report MIT-LCS-TM-632, MIT Laboratory for Computer Science,
  Dec. 2001.

\bibitem{Boulkenafed2003}
M.~Boulkenafed and V.~Issarny.
\newblock {AdHocFS: Sharing Files in WLANS}.
\newblock In {\em The Second IEEE International Symposium on Network Computing
  and Applications ({NCA'03})}, Apr. 2003.

\bibitem{Burrell2004}
J.~Burrell, T.~Brooke, and R.~Beckwith.
\newblock Vineyard computing: Sensor networks in agricultural production.
\newblock {\em Pervasive Computing, IEEE}, 3(1):38--45, Jan. 2004.

\bibitem{Cooley2004}
J.~Cooley, C.~Taylor, and A.~Peacock.
\newblock {ABS: The Apportioned Backup System}.
\newblock Technical Report 6.824, MIT Laboratory for Computer Science, 2004.

\bibitem{Courtes2006}
L.~Courtès, M.-O. Killijian, and D.~Powell.
\newblock {Storage Tradeoffs in a Collaborative Backup Service for Mobile
  Devices}.
\newblock Technical Report 05673, LAAS-CNRS, Apr. 2006.

\bibitem{Frenkiel2000}
R.~H. Frenkiel, B.~R. Badrinath, J.~Borràs, and R.~D. Yates.
\newblock {The Infostations Challenge: Balancing Cost and Ubiquity in
  Delivering Wireless Data}.
\newblock {\em {IEEE Personal Communications}}, 7(2):66--71, Apr. 2000.

\bibitem{Guy1998}
R.~Guy, P.~Reiher, D.~Ratner, M.~Gunter, W.~Ma, and G.~Popek.
\newblock {Rumor: Mobile Data Access Through Optimistic Peer-to-Peer
  Replication}.
\newblock In {\em The Workshop on Mobile Data Access}, Nov. 1998.

\bibitem{Mosaic2004a}
M.-O. Kilijian, D.~Powell, M.~Banâtre, P.~Couderc, and Y.~Roudier.
\newblock {Collaborative Backup for Dependable Mobile Applications}.
\newblock In {\em The 2nd International Workshop on Middleware for Pervasive
  and Ad-Hoc Computing (Middleware)}. ACM, Oct. 2004.

\bibitem{Mosaic2004}
M.-O. Kilijian, D.~Powell, M.~Banâtre, P.~Couderc, and Y.~Roudier.
\newblock {MoSAIC: Mobile System Availability Integrity and Confidentiality}.
\newblock Technical report, ACI SI, 2004.

\bibitem{Kubiatowicz2000}
J.~Kubiatowicz, D.~Bindel, Y.~Chen, S.~Czerwinski, P.~Eaton, D.~Geels,
  R.~Gummadi, S.~Rhea, H.~Weatherspoon, W.~Weimer, C.~Wells, and B.~Zhao.
\newblock {OceanStore: An Architecture for Global-Scale Persistent Storage}.
\newblock In {\em The Ninth International Conference on Architectural Support
  for Programming Languages and Operating Systems ({ASPLOS 2000})}, pages
  190--201, Nov. 2000.

\bibitem{Kumar1995}
P.~Kumar and M.~Satyanarayanan.
\newblock {Flexible and Safe Resolution of File Conflicts}.
\newblock In {\em {USENIX} Winter}, pages 95--106, 1995.

\bibitem{Loo2003}
B.~T. Loo, A.~LaMarca, and G.~Borriello.
\newblock {Peer-To-Peer Backup for Personnal Area Networks}.
\newblock Technical Report IRS-TR-02-015, Intel Research Seattle - University
  of California at Berkeley, May 2003.

\bibitem{Minsky2002}
Y.~Minsky and A.~Trachtenberg.
\newblock {Practical set reconciliation}.
\newblock Technical Report 2002-03., Boston University, 2002.

\bibitem{Patterson1988}
D.~A. Patterson, G.~Gibson, and R.~H. Katz.
\newblock {A Case for Redundant Arrays of Inexpensive Disks (RAID)}.
\newblock In {\em The International Conference on Management of Data (SIGMOD)},
  pages 109--116, June 1988.

\bibitem{Rabin1989}
M.~O. Rabin.
\newblock {Efficient dispersal of information for security, load balancing, and
  fault tolerance}.
\newblock {\em Journal of the ACM}, 36(2):335--348, Apr. 1989.

\bibitem{Ratner1999}
D.~Ratner, P.~Reiher, and G.~J. Pope1.
\newblock {Roam: A Scalable Replication System for Mobile Computing}.
\newblock In {\em The Workshop on Mobile Databases and Distributed Systems
  (MDDS)}, pages 96--104, Sept. 1999.

\bibitem{Satyanarayanan1990}
M.~Satyanarayanan.
\newblock {Scalable, Secure and Highly Available Distributed File Access}.
\newblock {\em IEEE Computer}, 23(5):9--21, May 1990.

\bibitem{Shah2003}
R.~C. Shah, S.~Roy, S.~Jain, and W.~Brunette.
\newblock Data mules: Modeling a three-tier architecture for spare sensor
  networks.
\newblock In {\em The First IEEE International Workshop on Sensor Network
  Protocols and Applications}, pages 30--41, May 2003.

\bibitem{Sobti2004}
S.~Sobti, N.~Garg, F.~Zheng, J.~Lai, Y.~Shao, C.~Zhang, E.~Ziskind,
  A.~Krishnamurthy, and R.~Y. Wang.
\newblock {Segank: A Distributed Mobile Storage System}.
\newblock In {\em The Third USENIX Conference on File and Storage
  Technologies}, Apr. 2004.

\bibitem{Srivastava2006}
M.~Srivastava, M.~Hansen, J.~Burke, A.~Parker, S.~Reddy, G.~Saurabh, M.~Allman,
  V.~Parxson, and D.~Estrin.
\newblock {Wireless Urban Sensing Systems}.
\newblock Technical Report~65, {Center for Embedded Networked Sensing Systems,
  UCLA}, Apr. 2006.

\bibitem{Terry1995}
D.~B. Terry, M.~M. Theimer, K.~Petersen, A.~J. Demers, M.~J. Spreitzer, and
  C.~H. Hauser.
\newblock {Managing Update Conflicts in {B}ayou, a Weakly Connected Replicated
  Storage System}.
\newblock In {\em The 15th {ACM} Symposium on Operating Systems Principles
  ({SOSP}'95)}, Dec. 1995.

\end{thebibliography}
